\documentstyle[twoside,fleqn,espcrc2,epsf,rotate]{article}

\def\Ps56{{\cal P}_Z(s_{56})}
\def\A60{{\cal A}^{{\rm tree}}_6}
\def\d{{\rm d}}

\newcommand\as{{\alpha_s}}

\newcommand\NLO{next-to-leading order }

\newcommand\qb{{\bar q}}
\newcommand\Qb{{\bar Q}}
\newcommand{\Nc}{{N_C}}
\newcommand{\NA}{{N_A}}
\newcommand{\Nf}{{N_f}}

\newcommand\Tr{{\rm Tr}}

\newcommand{\beq}{\begin{equation}}
\newcommand{\eeq}{\end{equation}}
\newcommand{\beqn}{\begin{eqnarray}}
\newcommand{\eeqn}{\end{eqnarray}}
\newcommand\nn{\nonumber}

\title{
Four-jet production in $e^+e^-$ annihilation at next-to-leading order
\thanks{This research was supported in part by the EEC Programme "Human Capital
and Mobility", Network "Physics at High Energy Colliders", contract
PECO ERBCIPDCT 94 0613 as well as by the Hungarian Scientific Research
Fund grant OTKA T-016613 and the Research Group in Physics of the
Hungarian Academy of Sciences, Debrecen.}}
\author{Zolt\'an Nagy
\address{Department of Theoretical Physics, KLTE, H-4010
Debrecen P.O.Box 5, Hungary}
and Zolt\'an Tr\'ocs\'anyi
\address{Institute of Nuclear Research of the Hungarian Academy of
Sciences, H-4001 Debrecen P.O.Box 51, Hungary}}

\begin{document}

\begin{titlepage}
\vspace*{-2cm}
\begin{flushright}
hep-ph/9708344 \\
\today \\
\end{flushright}
\vskip .5in
\begin{center}
{\large\bf
Four-jet production in $e^+e^-$ annihilation at next-to-leading order
\footnote{Presented by Z. Tr\'ocs\'anyi at the International Euroconference
on Quantum Chromodynamics ({\em QCD '97}), Montpellier, France,
July 3--9, 1997}}\\
\vskip 5mm 
Zolt\'an Nagy$^a$ and Zolt\'an Tr\'ocs\'anyi$^{b,a}$ \\
\vskip 2mm
\it $^a$Department of Theoretical Physics, KLTE, H-4010
Debrecen P.O.Box 5, Hungary \\
\it $^b$Institute of Nuclear Research of the Hungarian Academy of
Sciences, H-4001 Debrecen P.O.Box 51, Hungary \\
\end{center}
\vskip 5mm 

\begin{center} {\large \bf Abstract} \end{center}
\begin{quote}
We present a partonic Monte Carlo event generator that can be used
for calculating the group independent kinematical functions of any
infrared safe four-jet observable at \NLO accuracy. As an example, we
calculate the differential distribution of the $\Pi_1$ and $\Pi_4$
Fox-Wolfram moments. We find large K factors (K $>$ 2). The effect of
the radiative correction is to increase the overall normalization, but
not to reduce the renormalization scale dependence significantly. 
\end{quote}
\end{titlepage}
\setcounter{footnote}{0}

\newpage
\begin{abstract}
We present a partonic Monte Carlo event generator that can be used
for calculating the group independent kinematical functions of any
infrared safe four-jet observable at \NLO accuracy. As an example, we
calculate the differential distribution of the $\Pi_1$ and $\Pi_4$
Fox-Wolfram moments. We find large K factors (K $>$ 2). The effect of
the radiative correction is to increase the overall normalization, but 
not to reduce the renormalization scale dependence significantly.
\end{abstract}

\maketitle

\section{INTRODUCTION}

The analysis of four-jet production in $e^+e^-$ annihilation is an
important tool to learn about the the basic properties of Quantum
Chromodynamics (QCD), the theory of strong interactions.
So far the four-jet data were used mainly for color
factor measurements \cite{LEP}, because the theoretical prediction of
perturbative QCD has not been known at \NLO accuracy that is needed
for precision $\as$ measurements. In this contribution we report the
construction of a partonic event generator that can be used for
calculating the radiative corrections to the group independent
kinematical functions of any infrared safe four-jet observables in
electron positron annihilation. These results make possible the
simultaneous precision measurement of the strong coupling and the color
charge factors using LEP or SLC data. As an example, we present the
differential distributions of two event shape variables that are
non-trivial for four-jet like events --- the $\Pi_1$ and $\Pi_4$
Fox-Wolfram moments \cite{FWmoments}.

\section{THE METHOD}

The higher order correction to the leading order partonic cross section
is a sum of two terms, the real and virtual corrections:
\beq
\label{sNLO}
\sigma^{\rm NLO} \equiv \int \d\sigma^{\rm NLO}
= \int_5 \d\sigma^{\rm R} + \int_4 \d\sigma^{\rm V}\:,
\eeq
where in case of four-jet observables, the integrals are over the
five- and four-particle phase space respectively. These two integrals
are both divergent in $d=4$ space-time dimensions, however, their sum
is finite for infrared safe physical quantities. In recent years
several general methods have been developed for exposing the
cancellation of infrared divergences directly at the    
integrand level \cite{slicing,residue,CSdipole}. We use a modified
version of the dipole formalism of Catani and Seymour
\cite{CSdipole}.The formal result of this cancellation is
that the \NLO correction is a sum of two finite integrals,
\beq           
\label{sNLO2}
\sigma^{\rm NLO}
= \int_5 \d\sigma_5^{\rm NLO} + \int_4 \d\sigma_4^{\rm NLO}\:,
\eeq                    
where the first term is an integral over the available five-parton
phase space (as defined by the jet observable) and the second one is
an integral over the four-parton phase space.  The distinct feature of
this formalism as compared to other cancellation methods is that a
single subtraction term is used for the regularization of the real
cross section. This subtraction term provides a smooth approximation of
the real cross section in all of its singular limits (soft and collinear 
regions), resulting in a well-converging partonic Monte Carlo program.

The main ingredients of the calculation are the four-parton \NLO and
five-parton Born level squared matrix elements. The helicity amplitudes
from which the latter can be constructed have been known for almost a
decade \cite{5parton}. Recently, there was important development in the
calculation of the virtual corrections for the processes
$e^+e^-\to \qb q \Qb Q$ and $\qb q g g$. On one hand Campbell, Glover and
Miller made FORTRAN programs that calculates the \NLO corrections to
the four parton processes via the production of an s-channel virtual
photon publicly available \cite{CGM}, while on the other, the new
techniques developed by Bern, Dixon and Kosower in the calculation of
one-loop multiparton amplitudes \cite{BDKannals} made possible the
derivation of explicit analytic expressions for the helicity amplitudes
of the $e^+e^-\to Z^0,\,\gamma^*\to \qb q \Qb Q$ and $\qb q g g$ processes
\cite{BDKW4q,BDK2q2g}. These results ---  and so ours --- are valid in the
limit when all quark and lepton massess are set to zero.  We use the
helicity amplitudes of refs.~\cite{BDKW4q,BDK2q2g} for the loop corrections. 

\section{GENERAL STRUCTURE OF THE CROSS SECTION}

Once the integrations in eq.~(\ref{sNLO2}) are carried out, the \NLO
differential cross section for a four-jet observable $O_4$ takes the
form
\beq
\label{general}
\frac{O_4}{\sigma_0}\frac{\d \sigma}{\d O_4}(O_4)
= \left(\frac{\as(\mu) C_F}{2\pi}\right)^2 B_{O_4}(O_4)
\eeq 
\[
+ \left(\frac{\as(\mu) C_F}{2\pi}\right)^3
\left[B_{O_4}(O_4)\frac{\beta_0}{C_F}
 \ln\frac{\mu^2}{s} + C_{O_4}(O_4)\right]\:.
\]
In eq.~(\ref{general}) $\sigma_0$ denotes the Born cross section for the
process $e^+e^-\to \qb q$,
$\beta_0 = \left(\frac{11}{3}C_A - \frac{4}{3} T_R \Nf\right)$
with the normalization $T_R=1/2$ in ${\rm Tr}(T^aT^{\dag b})=T_R\delta^{ab}$,
$s$ is the total c.m.\ energy squared, $\mu$ is the renormalization scale, 
while $B_{O_4}$ and $C_{O_4}$ are scale independent functions,
$B_{O_4}$ is the Born approximation and $C_{O_4}$ is the radiative correction.
The Born approximation and the higher order correction are linear and
quadratic forms of ratios of eigenvalues of the Casimir operators of the
underlying gauge group \cite{NTloopamps}:
\beq
B_{O_4} = B_0 + B_x\,x + B_y\,y \:,
\eeq
and
\beq
C_{O_4} = C_0 +\,C_x\,x + C_y\,y + C_z\,z
\eeq
\[ \qquad
+\,C_{xx}\,x^2 + C_{xy}\,x\,y + C_{yy}\,y^2 \:.
\]
The $x$ and $y$ parameters are ratios of the quad\-ratic Casimirs,
$x=C_A/C_F$ and $y=T_R/C_F$, while $z$ is related to the square of a
cubic Casimir,
\beq
C_3 = \sum_{a,b,c=1}^\NA \Tr(T^a T^b T^{\dag c}) \Tr(T^{\dag c} T^b T^a)\:,
\eeq
via $z=\frac{C_3}{\Nc C_F^3}$.

\section{RESULTS}

Dixon and Signer obtained the first ever complete results for four-jet
observables at \NLO accuracy \cite{DSjets}, which were four-jet rates
for three different clustering algorithms: the Durham \cite{durham},
the Geneva \cite{geneva} and the E0 \cite{E0} schemes. In order to
compare the two programs, we have also calculated these quantities at
the same values of the parameters and found very good agreement
\cite{NTDpar}.


As a new result, we present the \NLO prediction for Fox-Wolfram moments
$\Pi_1$ and $\Pi_4$ \cite{FWmoments}. These observables were constructed
in such a way that they vanish for coplanar events, thus have non-trivial
value for four-jet like events. The variable $\Pi_1$ is defined as
\beq
\Pi_1 =
\sum_{i,j,k} \frac{|\vec{p}_i| |\vec{p}_j| |\vec{p}_k|}{\sqrt{s^3}}
(\hat{p}_i\times \hat{p}_j \cdot \hat{p}_k)^2\:,
\eeq
where $\vec{p}_i$ is the three-momentum of particle $i$, $\hat{p}_i$ is
the unit vector along the momentum $\vec{p}_i$, and
the sum runs over all final state particles in an event.
The variable $\Pi_4$ has more complicated definition:
\beqn
&&\Pi_4 =
\sum_{i,j,k} \frac{|\vec{p}_i| |\vec{p}_j| |\vec{p}_k|}{\sqrt{s^3}}
(\hat{p}_i\times \hat{p}_j \cdot \hat{p}_k)^2
\\ \nn &&\qquad
\left[(\hat{p}_i\cdot \hat{p}_j)^2
     +(\hat{p}_j\cdot \hat{p}_k)^2
     +(\hat{p}_k\cdot \hat{p}_i)^2\right]\:.
\eeqn

We present histograms for the various group independent kinematical
correction functions $C_i$ for $\Pi_1$ in Figure~1.  We find that the
correction functions $C_0$ and $C_x$ are large and positive, while
the functions $C_y$ and $C_{yy}$ are large and negative. We have not
shown the $C_z$ contribution, that is negligible.
\begin{figure}
\epsfxsize=7.3cm \epsfbox{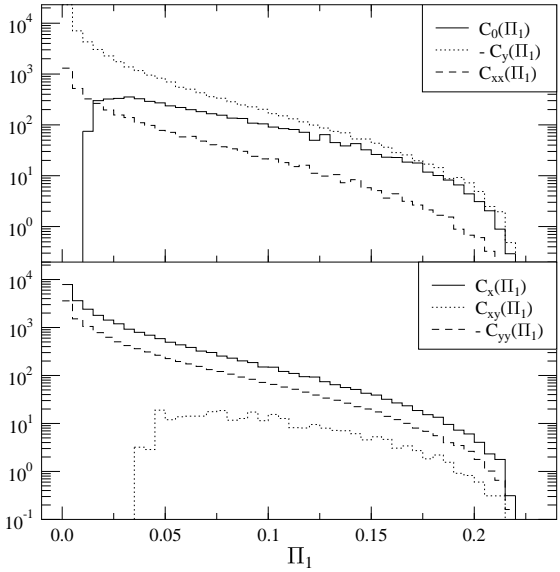} 
\vspace*{-16pt}
\caption
{Next-to-leading order corrections to the group independent kinematical
functions of the event shape variable $\Pi_1$.}
\end{figure}

We take SU(3) as underlying gauge group, and obtain the \NLO prediction
for the moments using the QCD values $x=9/4$ and $y=3/8$. The results ---
obtained for five light quark flavors at the $Z^0$ peak with
$M_Z = 91.187$\,GeV, $\Gamma_Z = 2.49$\,GeV, $\sin^2\theta_W=0.23$ and
$\as(M_Z) = 0.118$ --- are plotted in Figure~2 and 3. The light grey
bands indicate the renormalization scale dependence in the range
$0.1<x_\mu=\mu/\sqrt{s}<2$ at leading order, while the dark bands show
that at next-to-leading order. We see that the effect of the higher order
correction is to increase the overall normalization, but no significant
scale-dependence reduction occurs.
\begin{figure}
\vspace*{-23pt}
\epsfxsize=7.5cm
\epsfbox{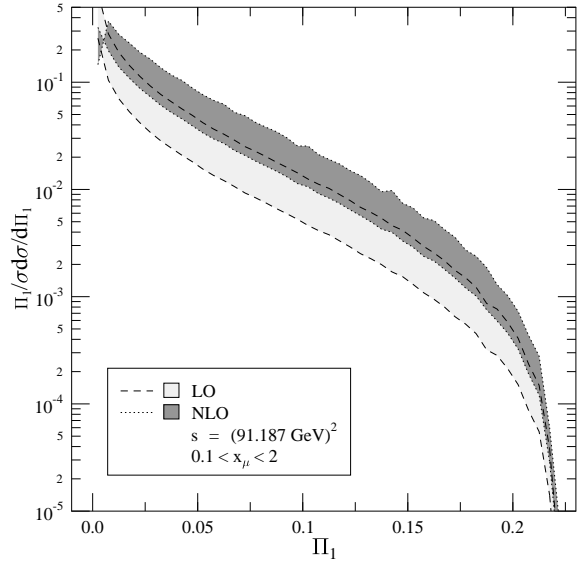}
\vspace*{-12pt}
\caption
{Next-to-leading order QCD prediction of the event shape variable $\Pi_1$.}
\end{figure}
\begin{figure}
\epsfxsize=7.5cm
\epsfbox{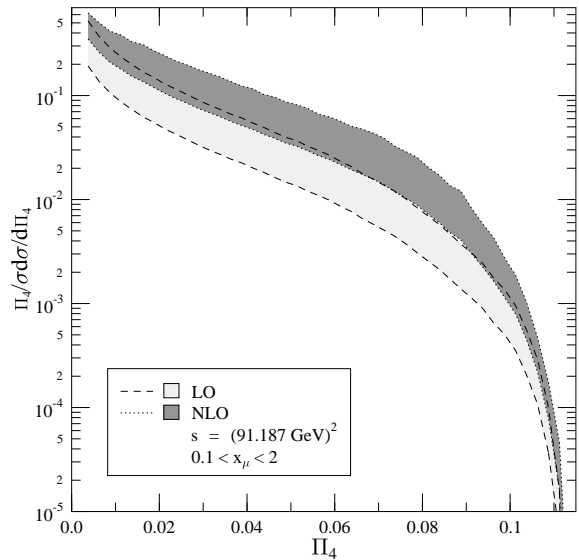} 
\vspace*{-16pt}
\caption
{Next-to-leading order QCD prediction of the event shape variable $\Pi_4$.}
\end{figure}

In order to see the renormalization scale dependence in more details,
we define the average value of these shape variables as
\beq                         
<O_4> = \frac{1}{\sigma} \int_0^1 \d O_4\,O_4\frac{\d \sigma}{\d O_4}\:,
\eeq           
and study the dependence of the average value of the $\Pi_1$ variable on
the scale in Figure~4. We see that there remains
substantial scale dependence at \NLO showing that the uncalculated
higher order corrections are presumably large. The feature is similar
for the $\Pi_4$ variable, but the residual scale dependence is even
larger.
\begin{figure}
\epsfxsize=5cm
\begin{rotate}[r]{\epsfbox{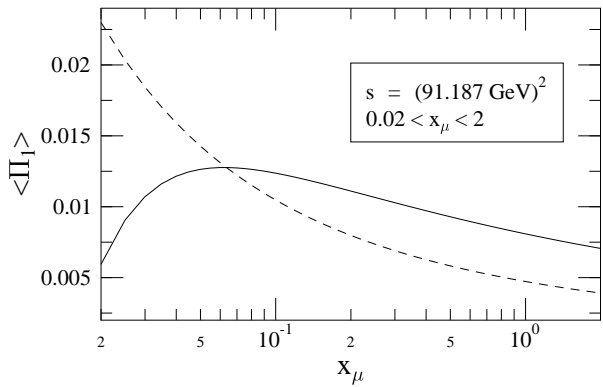}}\end{rotate}
\vspace*{-20pt}
\caption
{Renormalization scale dependence of the average value of the moment $\Pi_1$.}
\end{figure}

The same conclusion is drawn if we look at the dependence of the K
factors on the observables as depicted in Figure~5. In case of the
$\Pi_1$ parameter the K factor is slightly above two for the whole
range, while for $\Pi_4$ it is even larger.  This suggests that $\Pi_4$
cannot be reliable calculated in perturbation theory.
\begin{figure}
\epsfxsize=5.6cm
\begin{rotate}[r]{\epsfbox{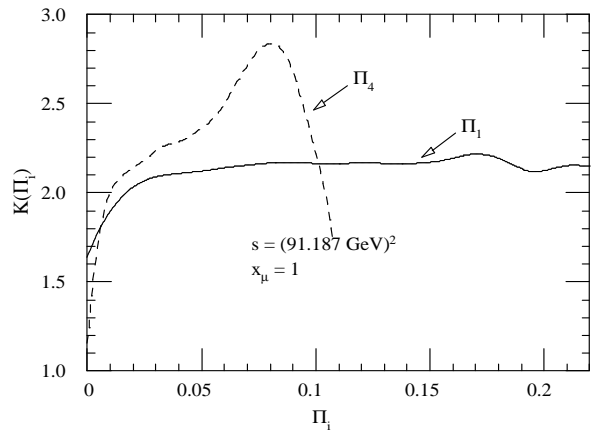}}\end{rotate}
\vspace*{-16pt}
\caption
{K factors of the $\Pi_1$ and $\Pi_4$ moments.}
\end{figure}

\section{CONCLUSION}

In this talk we presented for the first time a \NLO calculation of
the differential cross section of two four jet shape variables, the
$\Pi_1$ and $\Pi_4$ Fox-Wolfram moments.  We gave explicit results both
for the full \NLO cross sections and for the group independent
kinematical functions of the radiative corrections. The effect of the
radiative correction was to increase the overall normalization, but not
to reduce the renormalization scale dependence significantly. 

These results were produced by a partonic Monte Carlo program DEBRECEN
that can be used for the calculation of QCD radiative corrections to
the differential cross section of any kind of infrared safe four-jet
observable in electron-positron annihilation.

We thank L. Dixon for providing us the {\em Mathematica} files of the
one-loop four-parton helicity amplitudes.

\def\np#1#2#3  {Nucl.\ Phys.\ #1 (19#3) #2}
\def\pl#1#2#3  {Phys.\ Lett.\ #1 (19#3) #2}
\def\prep#1#2#3  {Phys.\ Rep.\ #1 (19#3) #2}
\def\prd#1#2#3 {Phys.\ Rev.\ D #1 (19#3) #2}
\def\prl#1#2#3 {Phys.\ Rev.\ Lett.\ #1 (19#3) #2}
\def\zpc#1#2#3  {Zeit.\ Phys.\ C #1 (19#3) #2}
\def\cmc#1#2#3  {Comp.\ Phys.\ Comm.\ #1 (19#3) #2}
\def\anr#1#2#3  {Ann.\ Rev.\ Nucl.\ Part.\ Sci.\ #1 (19#3) #2}


\section{DISCUSSIONS}

{\bf A.P. Contogouris}, Universtity of Athens \\
{\em To produce your K factors you need a jet algorithm; however, there
are more than one algorithms. Do I understand correctly that your K
factors much depend on it?} 

{\bf Z. Tr\'ocs\'anyi} \\
{\em The K factors do depend on the observable. This is also the case for
three-jet observables. The smaller K factor the more reliable the
quantity can be calculated in perturbation theory. From this point of
view the Durham jet clustering algorithm (K$\simeq$1.6) is a better
observable than the event shape variables.}

{\bf A.P. Contogouris} \\
{\em Suppose you compare some of your K factors with those for three-jet
production in $e^+e^-$ annihilation, corresponding to equivalent
observables (e.g.\ shape variable). How do your K factors compare? Also,
how do they compare with Drell-Yan lepton pari production, one of the
first processes for which K-factors were calculated?}

{\bf Z. Tr\'ocs\'anyi} \\
{\em The K factor for the C parameter for instance, is about 1.5, i.e.\
much smaller than in our case. The Drell-Yan K factor is 1.8--2,
which is again quite large, but somwhat smaller than our values.}

\end{document}